\documentclass[aps,twocolumn,amsmath,showpacs,floatfix]{revtex4}
\usepackage[dvips]{color}
\usepackage{psfig,newlfont}
\usepackage[english]{babel}
\usepackage{graphicx}
\usepackage[latin1]{inputenc}
\begin{document}

\title{Synchronization and Stability in Noisy Population Dynamics}

\author{Sabrina B.L. Araujo and M.A.M. de Aguiar}

\affiliation{Instituto de Física 'Gleb Wataghin',
Universidade Estadual de Campinas, \\
Caixa Postal 6165, 13083-970 Campinas, São Paulo, Brazil}

\begin{abstract}

We study the stability and synchronization of predator-prey
populations subjected to noise. The system is described by patches
of local populations coupled by migration and predation over a
neighborhood. When a single patch is considered, random
perturbations tend to destabilize the populations, leading to
extinction.  If the number of patches is small, stabilization in
the presence of noise is maintained at the expense of
synchronization. As the number of patches increases, both the
stability and the synchrony among patches increase. However, a
residual asynchrony, large compared with the noise amplitude, seems
to persist even in the limit of infinite number of patches.
Therefore, the mechanism of stabilization by asynchrony recently
proposed by R. Abta et. al.\cite{pabta}, combining noise, diffusion
{\it and nonlinearities}, seems to be more general than first
proposed.

\end{abstract}

%

\maketitle

The model proposed independently by Lotka \cite{lotka} and Volterra
\cite{volterra} was probably the first to describe mathematically the
dynamics of predators and preys. Its success and widespread use by
early biologists is mostly due to its ability to qualitatively
describe the population oscillations of both preys and predators
\cite{bmurray}. The model, however, is well known to be unstable under
the addition of noise, which causes the amplitude of the population
oscillations to increase until one (or both) species eventually
becomes extinct. Stability, in the sense of co-existence of both
species, can be regained if several patches of populations are coupled
via dispersal or predation. Computational simulations have shown that
stability increases (with respect to noise amplitude) with the number
of patches considered \cite{pdonalson,pwilson,pbriggs}. The ultimate
reason for the stabilization was recently pinned down by R. Abta et.
al.\cite{pabta}, who studied in detail the case of two patches. They
concluded that the crucial condition for stabilization is the
development of an asynchrony between the population oscillations in
each patch, resulting from the combined action of diffusion and
noise. Such an asynchrony develops if the frequency of the population
oscillations depends on their amplitude.

Although the results of ref.\cite{pabta} are of theoretical and
conceptual importance, simple Lotka-Volterra equations (LV, for
short) are seldom used to model population dynamics now-a-days.
Instead, models involving logistic type of interactions displaying
limit cycles or chaotic attractors have become common
\cite{pschaffer,php,phastings,psarkar}. In this report we discuss
the problem of stabilization and synchronization for a
predator-prey system displaying an attracting limit cycle
\cite{psabrina}. For this system all asymptotic orbits have the
same oscillation frequency, which is the frequency of the
attractor. As a consequence, orbits displaced from the attractor by
a perturbation will return to the attractor and to the same
frequency of oscillation, different from LV model. We shall call
this system LC (limit cycle) for short. The system is only weakly
stable in the presence of noise, in the sense that it becomes
unstable when the noise amplitude crosses a threshold which is very
small. Therefore, for noise amplitudes that are not too small, LC
behaves like LV, with noise driving one or both species to
extinction.

A spatial version of the LC model, SLC, can be constructed by
allowing patches of local populations to interact. If the patches
are strongly coupled, the dynamics in each spatial region
synchronizes and the system behaves like a single well mixed
population, identical to the original LC model. Here we study the
stability and synchronization of the spatial model under random
perturbations in the strongly coupled regime. Because of the
attracting limit cycle, it is not clear that an asynchrony among
patches will develop, since perturbed orbits will always have
similar amplitudes and, therefore, similar oscillation frequencies.
For a system with only two patches, we show that desynchronization
indeed takes place, leading to the stabilization of the population
oscillations, in agreement with the results of \cite{pabta} for the
LV model. As more patches are added, SLC system becomes stable
under larger noise amplitudes \cite{pdonalson} and the asynchrony
decreases exponentially with system size. However, a residual
asynchrony, much larger than the noise amplitude, seems to survive
even in the limit of infinite number of patches. Therefore, even
for large systems displaying an attracting limit cycle, the
combined action of diffusion and noise still plays a crucial role
in desynchronizing the patches.

The LC predator-prey model with noise is given by the equations
\begin{equation}
\begin {array}{l}
\displaystyle{x_{n+1}=\left[
\frac{x_{n}}{x_{n}(1-a)+a}\right]
P_{x}(y_{n}) +\eta_x},\\ \\
\displaystyle{y_{n+1}=y_{n}[e^{-d_{1}}+F_{y}(x_{n})]+\eta_y}
\end{array}
\label{eq1}
\end{equation}
where $P_{x}(y)=e^{-y/\alpha}$ accounts for the predation of $y$
upon the prey $x$ and $F_{y}(x)=1-e^{-x/\beta}$ for the
reproduction of $y$ due to the feeding upon $x$. The variables
$\eta_x$ and $\eta_y$ are random numbers representing the external
noise whose distributions are homogeneous and limited to the
interval $[-\Delta,\Delta]$.

To understand the role of each term in these equations let us first
consider $\eta_x=\eta_y=0$. Then, in the absence of predators,
$P_x(0)=1$ and the population of preys converges to the normalized
value $x=1$, provided $a<1$. As the number of predators increases,
$P_x(y)$ decreases, reducing the population of preys. Similarly, in
the absence of preys $F_y(0)=0$ and the population of predators
decreases steadily because of the intrinsic death rate $d_1$. When
the number of preys is sufficiently large so that $e^{-d_1}+F_y(x)
> 1$ the number of births becomes larger than the number of deaths
and the population of predators grows.

Following \cite{psabrina} we fix the model parameters at
$a=e^{-1}$, $\alpha=0.8$ and $\beta=0.4$. In the absence of noise
the system displays two kinds of attractive orbits, depending on
the value of the death rate $d_1$: for $0<d_1<0.6$ the attractor is
a limit cycle and for $d_1>0.6$ it is a fixed point. When the noise
is turned on it might happen that the population densities $x$ or
$y$ become less than zero. In this case we set it back to zero to
avoid negative values.

Throughout this paper we have fixed the death rate of the predator at
$d_1=0.1$. In this case the trajectories converge to a limit cycle
where the prey population oscillates between $2.7 \times 10^{-4}$ and
$0.3$. This attractor is only very weakly stable, since the addition
of noise with amplitudes of the order of $\Delta=6\times10^{-5}$ is
enough to drive both species to extinction.

To consider more than one patch we extend the LC model to
\cite{psabrina}:
\begin{eqnarray}
 x_{n+1}^{i,j}&=&\left[ \frac{x_{n}^{i,j}}{x_{n}^{i,j}
(1-a)+a}\right]\left<P_{x}(y_{n})\right>_{R} \nonumber\\
&&+\frac{m_{x}}{4}\left(\sum_{l,m}  x^{i+l,j+m}\right)-
m_{x}x^{i,j}+\eta_{x_{i,j}}\nonumber
\\\nonumber\\
y_{n+1}^{i,j}&=&y_{n}^{i,j}\left[e^{-d_{1}}+F_{y}
(\left<x_{n}\right>_{})\right]\nonumber\\
&&+\frac{m_{y}}{4}\left(\sum_{l,m}  y^{i+l,j+m}\right)-
m_{y}y^{i,j}+\eta_{y_{i,j}} \label{eq2}
\end{eqnarray}
where $i,j$ label the position of the patches on a two-dimensional
grid of $N\times N$ patches. This spatial version (SLC) allows
migration of predators and preys, with rates $m_y$ and $m_x$
respectively, and predation over a {\it predation neighborhood}
$R$. Here we consider square neighborhoods with sides $2R+1$
centered on the predator.

The sum over $l$ and $m$ in the migration terms is restricted only
to the four nearest neighbors of the site $(i,j)$. The averages
$\left<P_{x}(y_{n})\right>_{R}$ are performed over the patches that
are in predation neighborhood: $\left<P_{x}^{i,j}(y)\right>_{R} =
\frac{1}{N_{R}}\sum_{l,m=-R}^{R}P_{x}(y^{i+l,j+m})$ where
$N_R=(2R+1)^2$ is the number of patches within the neighborhood.
The feeding function $F_{y}(\left<x\right>_{R^{i,j}})$ is
calculated over the average number of $x$ on the predation
neighborhood: $F_{y}(\left<x\right>_{R}) = F_{y}
\left(\frac{1}{N_{R}} \sum_{l,m=-R}^{R}x^{i+l,j+m} \right)$. We fix
the migration rate of the prey and the predator at $m_x=0.01$ and
$m_y=0.1$ respectively, so that the migration rate of the predator
is larger then that of the prey \cite{pbascompte,bsole}.

In the SLC model, the average dynamic behavior can be very different
from that of the LC model. However, if the predation neighborhood
encompasses most of the patches in the grid, the system becomes
strongly coupled and the time evolution in each patch synchronizes
with the others\cite{psabrina}. \\

{\bf Two patches}\\

\begin{figure}[!ht]
    \centering{
    \includegraphics[width=7cm]{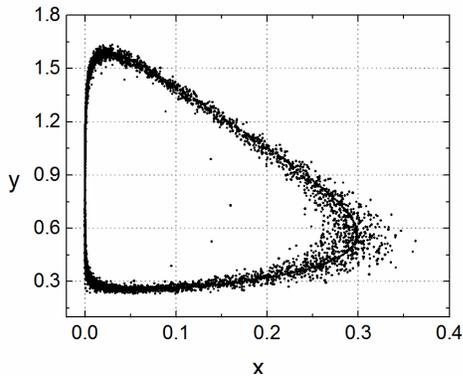}}
\caption{Phase-space trajectories for the case of two patches. The
plots show the time steps between 1,000 and 10,000 for only one
patch. The patches are synchronized up to 2,000 steps, when noise is
added and synchronization breaks down. The patches nearly synchronize
when $x\approx 0$ and get significantly desynchronized for large $x$.}
\label{fig1}
\end{figure}

In our first analysis we consider only two patches. In this case
Eq.(\ref{eq2}) reduces to
\begin{eqnarray}
 x_{n+1}^{i}&=&\left[ \frac{x_{n}^{i}}{x_{n}^{i}
(1-a)+a}\right]\left<P_{x}(y_{n})\right> \nonumber\\
&& + m_x x^{j}-m_{x}x^i+\eta_{x_i} \nonumber
\\\nonumber\\
y_{n+1}^{i}&=&y_{n}^{i}\left[e^{-d_{1}}+F_{y}
(\left<x_{n}\right>)\right]+m_y y^{j}\nonumber\\
&&-m_{y}y^{i}+\eta_{y_i}. \label{eq3}
\end{eqnarray}
where $i$ labels one of the patches and $j$ the other one. The
averages are given by $\left<P_{x}(y_{n})\right> =
\frac{1}{2}(P_x(y^i)+P_x(y^j))$ and $F_{y}(\left<x_{n}\right>) =
F_{y}(\frac{x^i+x^j}{2})$. In the absence of noise and with random
initial conditions, the time evolution of the populations synchronize
perfectly. However, when noise is added to the system, the populations
oscillate between synchronized and desynchronized phases. In our
simulations we let the system evolve without noise for 2,000 time
steps, which is enough to synchronize the patches.  Noise with
amplitude of $\Delta=6\times10^{-5}$ (which is enough to destabilize
the dynamics on a single patch) is then added to the system, which is
further evolved for another 28,000 steps.  Figure~\ref{fig1} shows the
populations in one of the patches in the phase-space for times between
1,000 and 10,000. The coupling between the patches stabilizes the
system. For larger noise amplitudes, of the order of
$\Delta=2\times10^{-4}$, instability sets in again and both species
may go extinct.

\begin{figure}[!ht]
    \centering{
    \includegraphics[height=6cm]{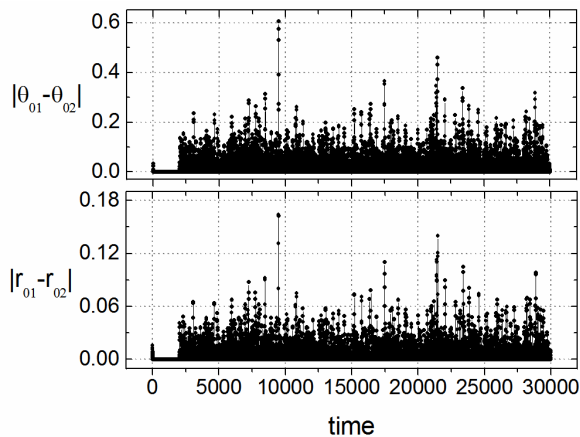}}
\caption{Phase (in radians) and amplitude differences between the
two patches for the trajectories shown in Figure~(\ref{fig1}). The
crests correspond to regions where $x$ is large. }
\label{fig2}
\end{figure}

In order to quantify the synchrony between the patches, we fixed a
reference point at $\vec{r_0}=(0.04,0.75)$ and measured the phase and
amplitude differences between the vectors $\vec{r}_{01} =
\vec{r_0}-\vec{r}_1 \equiv r_{01} e^{i\theta_{01}}$ and $\vec{r}_{02}
= \vec{r_0}-\vec{r}_2 \equiv r_{02} e^{i\theta_{02}} $, where
$\vec{r}_i$ is the phase space position of the populations in patch
$i$. Figure~\ref{fig2} shows the phase difference
$|\theta_{01}-\theta_{02}|$ and amplitude difference
$|r_{01}-r_{02}|$. As the initial conditions for each patch are
different, the phase and amplitude differences are initially
non-zero. However, the patches quickly synchronize in the absence of
noise (first 2,000 time steps) and de-synchronize again when noise is
added.\\

{\bf Increasing the number of patches}\\

When more patches are taken into account, the stability properties
change qualitatively. We considered the spatial model on a set of
$N \times N$ patches, Eq. (\ref{eq2}), with periodic boundary
conditions. As before, we fixed $d_1=0.1$, $\alpha=0.8$,
$\alpha=0.4$, $m_x=0.01$, $m_y=0.1$ and $\Delta=6\times10^{-5}$.
The only free parameters are the sizes of the grid, $N$, and the
predation radius, $R$. As in the case of two patches, it is
possible to synchronize the patches in the absence of noise if the
coupling is sufficiently strong \cite{psabrina}.

In order to study the effect of noise on large synchronized systems
we did simulations in which $N$ was varied but the ratio
$(2R+1)/N$, was kept constant. We considered 8 combinations of $N$
and $R$ with $(2R+1)/N= 5/6$: (1) $N=6$, $R=2$; (2) $N=18$, $R=7$;
(3) $N=30$, $R=12$; (4) $N=42$, $R=17$; (5) $N=54$, $R=22$; (6)
$N=66$, $R=27$; (7) $N=78$, $R=32$; (8) $N=90$, $R=37$. In all
cases the patches synchronized in the absence of noise.

The simulations started with random initial conditions and were
iterated by 60,000 time steps. Noise was added only after the first
5,000 steps. As in the case of two patches, we fixed the reference
point $\vec{r_0}=(0.04,0.75)$ and calculated the phase and
amplitude differences for each patch with respect to patch number
one. The temporal averages of the phase and amplitude differences
were computed for each individual patch for the last 50,000 time
steps, $\left<|\theta_1-\theta_i|\right>$ and
$\left<|\vec{r}_1-\vec{r}_i|\right>$ and also the global average,
$\Delta \bar{\theta} =
\frac{1}{N^2-1}\sum_{i=2}^{N^2}\left<|\theta_1-\theta_i|\right>$
and $\Delta \bar{r}=
\frac{1}{N^2-1}\sum_{i=2}^{N^2}\left<|\vec{r}_1-\vec{r}_i|\right>$.

In agreement with D. D.  Donalson et. al. \cite{pdonalson} we found
that the populations resist to higher noise amplitudes as the
number of patches increases. Moreover, for fixed noise amplitude
and coupling ratio, the patches tend to become more synchronized as
$N$ increases. Therefore, for large systems, stability and
asynchrony are not as correlated as in the case of two patches
\cite{pabta}.

\begin{figure}[!ht]
    \centering{
    \includegraphics[width=6cm]{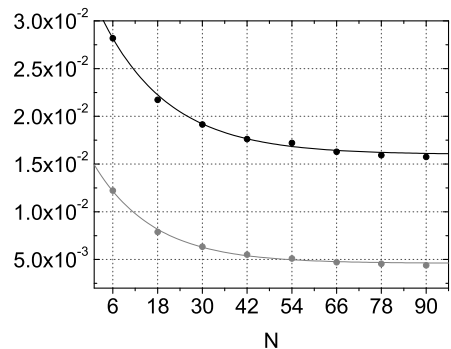}}
\caption{Global average of phase difference $\Delta\bar{\theta}$
(black), and amplitude difference $\Delta\bar{r}$ (gray), as a
function of grid size, $N$. The lines are first order exponential
decay fit.} \label{fig3}
\end{figure}

Figure~\ref{fig3} shows that the average asynchrony decreases
exponentially with grid size, following approximately the curve
$y(N)=A exp(-N/b)+c$. For the amplitude difference (gray curve) we
found $A=0.017$, $b=18$ and $c=0.016$ and for the phase difference
$A=0.011$, $b=16$ and $c=0.0046$. In the limit of infinitely many
patches the average amplitude difference tends to $c=0.046$, which
is significantly larger than the noise, $\Delta=6\times10^{-5}$.
The conclusion of these numerical experiments is that the
simultaneous presence of noise and diffusion seems to lead to
significant asynchrony even if the unperturbed dynamics has an
attractive limit cycle. For small systems this asynchrony can
stabilize the populations by allowing the migration of individuals
from more populated patches to those where extinction is imminent.
For large systems, asynchrony decreases exponentially fast with system
size but never disappears completely.

The mechanism responsible for the desynchronization in systems with
attractors seems to be the same proposed in \cite{pabta}, i.e.,
motion with amplitude depend frequencies. Since the dynamics tend
to bring perturbed orbits back to the attractor, a typical
trajectory always wanders in the vicinity of the attractor.
However, the nonlinear character of the equations amplify these
small deviations producing significant frequency differences that
are reflected in the desynchronization. Therefore, the mechanism of
R. Abta et. al.\cite{pabta}, combining noise, diffusion {\it and
nonlinearities}, seems to be more general than first proposed.\\

\noindent {\bf Acknowledgements} This work was partially supported
by Fapesp and CNPq.


\end{document}